\documentclass[fleqn,10pt]{wlscirep}
\usepackage[utf8]{inputenc}
\usepackage[T1]{fontenc}
\title{Diffracting and Non-diffracting random fields}
\author[1]{Patnala Vanitha}
\author[1]{Bhargavi M}
\author[2,*]{Venkateswarlu Annapureddy}
\author[1,**]{Gangi Reddy Salla}
\author[3,4]{Yoko Miyamoto }
\author[5]{R. P. Singh}

\affil[1]{Department of Physics, SRM University-AP,  Andhra Pradesh, India - 522240.}
\affil[2]{Department of Physics, National Institute of Technology  Tiruchirappalli, Tamil Nadu 620015, India.}
\affil[3]{Department of Engineering Science, The University of Electro-Communications, Chofugaoka, Chofu, Tokyo 1828585, Japan.}
\affil[4]{Institute for Advanced Science, The University of Electro-Communications, Chofugaoka, Chofu, Tokyo 1828585, Japan.}
\affil[5]{Physical Research laboratory, Navarangpura, Ahmedabad, India-380009.}
\affil[**]{Corresponding author: gangireddy.s@srmap.edu.in}

\begin{abstract}
We have generated and propagated both diffracting and non-diffracting speckles using the scattering of perfect optical vortices. The diffracting speckles have been realized in the near field and non-diffracting speckles have been realized in the far field, that is after taking the Fourier transform of near-field speckles using a simple convex lens. The exact analytical expressions have been provided for the size of both diffracting as well as non-diffracting speckles and compared with our experimental results. We found that the experimental results are in good agreement with the theoretical results. These results may find applications in classical cryptography and communication as we have both varying and non-varying random field patterns with propagation distance.
\end{abstract}
\begin{document}

\flushbottom
\maketitle
%
%
\thispagestyle{empty}

\section{Introduction}

Optical vortices have helical wave fronts due to their azimuthal phase along with a singularity at the center \cite{heckenberg1992generation}. They carry an orbital angular momentum (OAM) of $m\hbar$ per photon, $m$ being the order of the vortex \cite{torres2011twisted, franke2008advances}. These beams have attracted a lot interest due to their applications in science and technology \cite{molina2007twisted, yao2011orbital}. Recently, a new class of vortex beams have been introduced, namely perfect optical vortices (POV), which have order independent intensity distributions \cite{reddy2013experimental}. The POV beams can be generated using various techniques such as the Fourier transform of Bessel-Gauss (BG) beams \cite{ostrovsky2013generation, chen2013dynamics, vaity2015perfect, gori1987bessel}. Here, we have utilized the scattering of the POV beams for generating the diffracting as well as non-diffracting random patterns.

Speckle is a random granular light pattern and formed when a rough surface such as ground glass plate is illuminated by a coherent light beam \cite{reddy2014higher, hu2020does}. This random pattern is due to the interference of scattered wave fronts arriving from a large number of inhomogeneities present in the rough surface \cite{goodman2007speckle, dainty2013laser, franccon2012laser, jacquot2012interferometry}. These random patterns have significant applications in science and technology such as communication and cryptography \cite{ricklin2002atmospheric, schouten2003phase, kumar2019image, heeman2019application, sirohi1999speckle, cheng2007simplified, fercher1981flow, kermisch1975partially}. It has been observed that when optical vortices are scattered through a rough surface, the speckle size decreases with the increase in order \cite{reddy2014higher}. 

Since POV beams have order independent intensity distributions, it is of interest whether they produce order dependent speckles when scattered. We have previously scattered POV beams and found that the resulting speckle size, defined as the spatial length up to which the field is correlated, is independent of the order of the POV beams \cite{reddy2016non}. We have further studied these speckles in the far field, i.e., through a Fourier transforming lens, and found that they were non-diffracting \cite{cottrell2007nondiffracting, dudley2012controlling}. The size of the speckles can be controlled by changing the axicon parameter used in the generation of POV beams. We have also reported the generation of Bessel-Gauss coherence functions using the speckles obtained by scattering POV beams \cite{vanitha2021correlations}. While the speckle size in the far field is order independent, the speckle pattern itself is altered by the vortex, leading to specific cross-correlation functions dependent on the combination of POV orders.

In this work we examine in detail the change of speckle size during propagation and confirm that both diverging as well as non-diverging speckles can be generated from POV beams. As before the far field speckles are obtained by taking the Fourier transform of near field speckles using a plano-convex spherical lens. We derive the exact analytical expressions for the size of both near and far field speckles using the auto-correlation function. The size of near field speckles varies linearly with propagation distance and the size of far field speckles is independent of the propagation distance. We also verify our theoretical results experimentally for the near field and far field speckles.

\section{Theoretical Analysis}

Let us consider the field distribution of a perfect optical vortex (of order $m$) beam, which can be expressed as \cite{ostrovsky2013generation} 
\begin{equation}
E(\rho,\theta)= \delta\left(\rho-\rho_0\right)\ e^{im\theta}
\label{filed}
\end{equation}
where $\rho_0$ is the radius of the POV beam, and $\delta$ indicates the Dirac delta function. To allow for the POV beams having finite width $\varepsilon$, we replace the delta function with a narrow-Gaussian function $g\left(\rho-\rho_0; \varepsilon\right)$ so that
\begin{equation}
E(\rho,\theta)= g\left(\rho-\rho_0; \varepsilon\right)e^{im\theta}
\label{annularbeam}
\end{equation} 
In order to realize the diffracting and non-diffracting speckles, we propagate the POV beams through a rough surface, such as a ground glass plate (GGP). The theoretical formulation for the scattering of light beams through a GGP can be done using a random phase function $e^{i\Phi}$, where $\Phi$ is a random phase that varies in the range  0 to 2$\pi$. Now, the scattered field distribution is given by \cite{goodman2007speckle}
\begin{equation}
U(\rho,\theta)=  e^{i\Phi(\rho,\theta)}E(\rho,\theta)
\label{specklepattern}
\end{equation}
The inhomogeneities present in a ground glass plate are independent of each other and their size is of the order of 20 $\mu m$, which is very small compared to typical beam size. This implies that the auto-correlation of the phase exponential factor present in the above equation \ref{specklepattern} is provided by a Dirac-delta function as given below:
\begin{equation}
\langle e^{i[\Phi(\rho_1,\theta_1)-\Phi(\rho_2,\theta_2)]}\rangle=\delta\left(\rho_1-\rho_2\right)\delta\left(\theta_1-\theta_2\right)
\label{exponential}
\end{equation}
where $\langle a \rangle $ denote the ensemble average operation on $a$ \cite{goodman2015statistical}. 

The speckle size can be determined using the width of the auto-correlation function of a given speckle pattern. The autocorrelation function of a scattered field at a given plane ($z$) is defined as \cite{goodman2007speckle} 
\begin{equation}
\Gamma\left(r_1, \varphi_1;r_2,\varphi_2\right)=\langle U_1(r_1,\varphi_1,z)U_2^*(r_2,\varphi_2,z)\rangle
\end{equation}
For finding the speckle size for different propagation distances, we evaluate the above equation using Fresnel diffraction integral, which can be written in cylindrical coordinates as \cite{acevedo2018non},
\begin{eqnarray}
\Gamma_{12}\left( \Delta r\right)=\frac{e^{\frac{ik}{2z}\left( r_1^2-r_2^2\right) } }{\lambda^2 z^2}\int\int     \vert U_1(\rho,\theta)\vert^{2}
e^{-\frac{ik}{z}\left( \rho\Delta rcos(\varphi_s-\theta)
\right)}\rho d\rho d\theta
\label{gamma}
\end{eqnarray}
where
\begin{eqnarray}
\Delta rcos(\varphi_s-\theta)&=&\left[\left( r_1 cos\left( \varphi_1\right) -r_2 cos\left( \varphi_2\right)\right)cos\theta\right]            
 +  \left[\left( r_1 sin\left( \varphi_1\right) -r_2 sin\left( \varphi_2\right)\right)sin\theta\right]
\end{eqnarray}
and $\Delta r^2=r_1^2+r_2^2-2r_1r_2 cos\left(\varphi_2-\varphi_1\right) $. Note that equation \ref{exponential} has already been applied.

By using Eq. \ref{specklepattern}, we get the initial intensity as \cite{vanitha2021correlations}
\begin{eqnarray}
 U(\rho,\theta)U^*(\rho,\theta)&=& E(\rho,\theta)e^{i\Phi(\rho,\theta)} 
E^*(\rho,\theta) e^{-i\Phi(\rho,\theta)} \nonumber \\
&=& E(\rho,\theta)E^*(\rho,\theta)
\label{autocorrelation}
\end{eqnarray}
Using Eq. \ref{annularbeam} we obtain
\begin{eqnarray}
E(\rho,\theta)E^*(\rho,\theta)= g\left(\rho-\rho_0; \varepsilon\right)g\left(\rho-\rho_0; \varepsilon\right)  = g^2\left(\rho-\rho_0; \varepsilon\right)
\label{POV}
\end{eqnarray}
Under the approximation $\varepsilon\longrightarrow0$, one can replace $g^2\left(\rho-\rho_0; \varepsilon\right)$ with a single Dirac-delta function $\delta(\rho-\rho_0)$. By substituting this along with Eqs. \ref{autocorrelation} and \ref{POV} in Eq.\ref{gamma}, we get the auto-correlation function as
\begin{eqnarray}
\Gamma_{12}\left( \Delta r\right)=\frac{e^{\frac{ik}{2z}\left( r_1^2-r_2^2\right) } }{\lambda^2 z^2}\int\int  \delta(\rho-\rho_0)
e^{-\frac{ik}{z}\left( \rho\Delta rcos(\varphi_s-\theta)
\right)}\rho d\rho d\theta
\end{eqnarray}
Using Anger-Jacobi identity and the integral properties of Dirac-delta function, we obtain the correlation function as \cite{zwillinger2007table}

\begin{equation}  
\Gamma_{12}\left( \Delta r\right)=\frac{2\pi\rho_0 e^{\frac{ik}{2z}\left( r_1^2-r_2^2\right) } }{\lambda^2  z^2}J_0\left( \frac{k\rho_0}{z}\Delta r\right)
\end{equation} 
The above equation contains the zero-order Bessel function of the first kind \cite{vanitha2021correlations}. The speckle size is defined as the spatial length up to which correlations exist in the field. Here, we consider the speckle size as the distance of first zero of correlation function from center. The first zero of zeroth order Bessel function function $J_0\left( x\right) = 0$ happens at $x = 2.4$ and the correlation length or speckle size can be obtained as

\begin{equation} 
\Delta r=\frac{xz}{k\rho_0}=\frac{2.4z}{k\rho_0}
\label{near field}
\end{equation} 

It is clear from the above equation that the size of near-field speckles varies linearly with propagation distance $z$, is independent of order $m$, and inversely proportional to the ring radius $\rho_0$.

The far-field speckle pattern can be obtained by taking the Fourier transform of near-field speckle pattern, which can be realized in the lab using a simple convex lens (L2) of focal length $f_2$. The auto-correlation function $\Gamma_{12}^{'}\left( \Delta r^{'}\right) $ of the far field is  \cite{acevedo2018non} 
\begin{eqnarray}
\Gamma_{12}^{'}\left( \Delta r^{'}\right)=\frac{1}{\lambda^{2}f_2^{2}}\int\int    \vert U(\rho,\theta)\vert^{2}
e^{-\frac{ik}{f_{2}}\left( \rho\Delta r^{'}cos(\varphi_s-\theta)\right)}\rho d\rho d\theta
\label{gamma1}
\end{eqnarray} 
where $\Delta r^{'2}=r_1^{'2}+r_2^{'2}-2r_1^{'}r_2^{'} cos\left(\varphi_2^{'}-\varphi_1^{'}\right)$. 
By substituting the Eqs. \ref{autocorrelation} and \ref{POV} in Eq.\ref{gamma1}, then we get
\begin{eqnarray}
\Gamma_{12}^{'}\left( \Delta r^{'}\right)=\frac{1}{\lambda^{2}f_2^{2}}\int\int  \delta(\rho-\rho_0) 
e^{-\frac{ik}{f_{2}}\left( \rho\Delta r^{'}cos(\varphi_s-\theta)\right)}\rho d\rho d\theta
\end{eqnarray}
Once again the Anger-Jacobi identity  and the integral properties of Dirac-delta function give

\begin{equation}
\Gamma_{12}^{'}\left( \Delta r^{'}\right)= \frac{2\pi\rho_0}{\lambda^{2} f_{2}^2}J_{0}\left( \frac{k \rho_{0}}{f_{2}}\Delta r^{'}\right)  
\end{equation}
From the above equation, it is clear that the correlation function is independent of order $m$ as well as propagation distance $z$. The speckle size
\begin{equation} 
\Delta r^{'}=\frac{2.4f_2}{k\rho_0}
\label{far field}
\end{equation} 
is independent of propagation distance $z$, directly proportional to focal length $f_2$, and inversely proportional to ring radius $\rho_0$. Thus, we can say that the near field speckles are diffracting and the far field speckles are non-diffracting in nature. 

\section{Experimental Setup}

\begin{center}
   \begin{figure}[htb]
   \includegraphics[width=7.0cm]{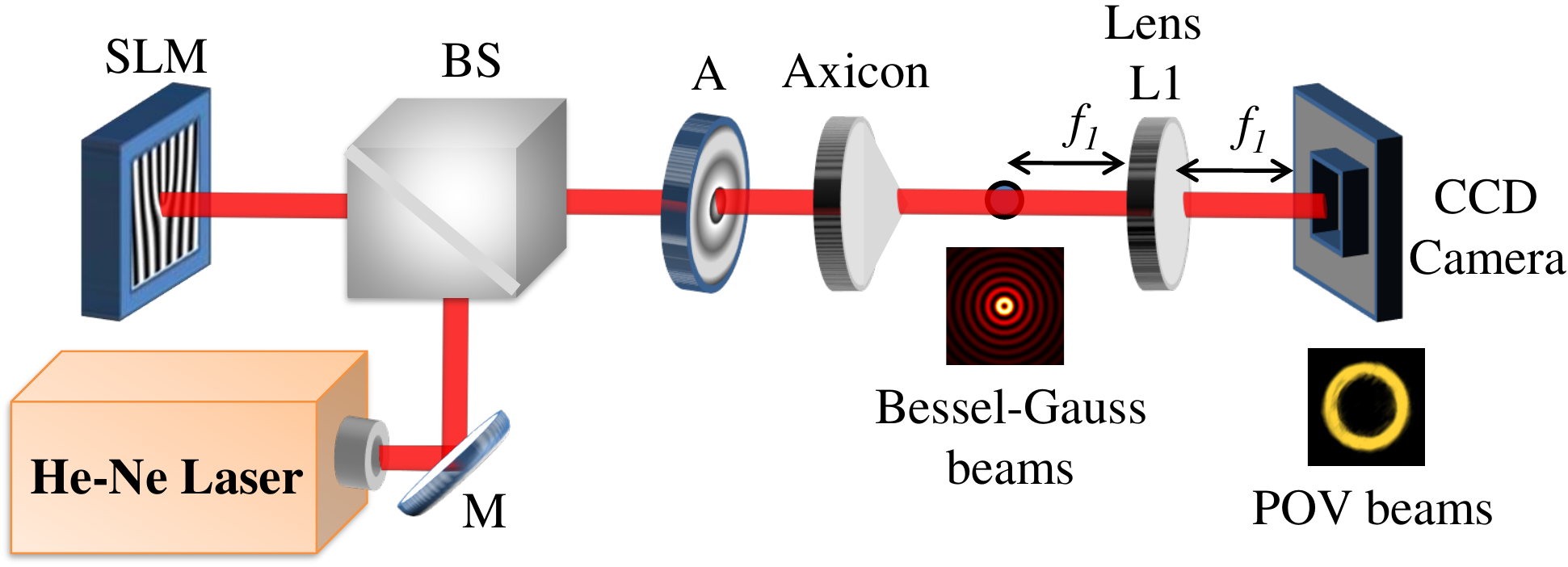}
    \caption{\textit{(Colour online) Experimental set up for the generation of POV beams using the Fourier transform of Bessel-Gauss beams (POV generating FT), Where SLM-Spatial Light Modulator, BS-Beam Splitter, A-Aperture, M-Mirror, $f_1$-focal length of L1.}}
    \label{fig:expt}
   \end{figure}
   \end{center}
The experimental setup designed to generate diffracting and non-diffracting speckles is shown in  Fig. \ref{fig:expt}. We have adopted the method of taking Fourier transform (FT) of Bessel-Gauss (BG) beams to produce the POV beams. Optical vortex beams are initially generated by illuminating a computer generated hologram displayed on a spatial light modulator. The beams then propagate through an axicon of apex angle $178^{o}$  \cite{bezerra2020sorting, alves2016using} to produce BG beams. The Fourier transform of BG beams is taken using a lens of focal length 30 cm, and at the back focal plane we obtain the POV beams (POV generating FT). We use a He-Ne laser of power 5 mW and wavelength 632.8 nm.

   \begin{center}
   \begin{figure}[htb]
   \includegraphics[width=7.0cm]{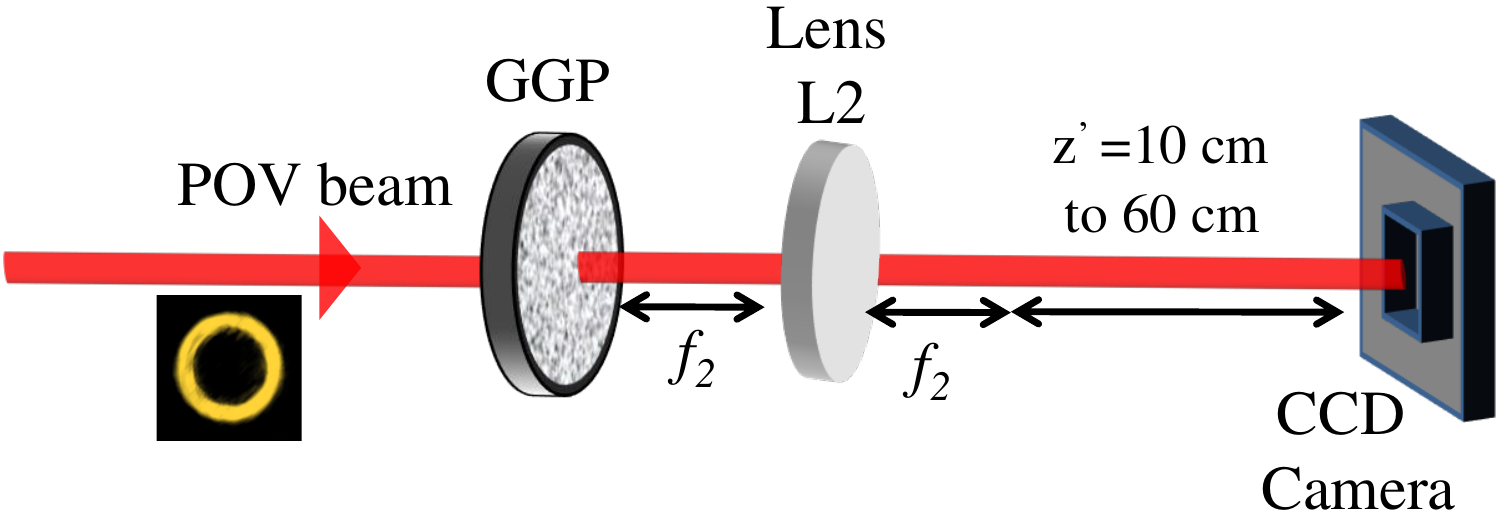}
    \caption{\textit{(Colour online) Experimental set up for producing the diverging and non-diverging speckle patterns, Where GGP-Ground Glass Plate, $f_2$-focal length of L2 used for speckle FT.}}
    \label{fig:expt1}
   \end{figure}
   \end{center}  

The generated POV beams are scattered through a ground glass plate (DG10-600, from Thorlabs) as shown in Fig. \ref{fig:expt1}. The ground glass plate is placed at the back focal plane of the lens L1 in Fig. \ref{fig:expt}, i.e. at the origin of POV beams. The near field speckles, without using the lens (L2) of focal length $f_2$ shown in Fig.\ref{fig:expt1}, are recorded just after the GGP. The diverging nature can be verified by looking at the size of the speckles at various distances within the range 20-70 cm from GGP at an interval of 5 cm. For generating the far field speckles, we take the Fourier transform of near field speckles using the lens (L2) placed at a distance of $f_2 = 50 cm$ from the GGP (speckle FT). For recording the POV beams and the corresponding speckles, we use a CCD camera (FLIR) with pixel size 3.14$\mu m$.

\section{Results \& Analysis}

\begin{center}
   \begin{figure}[htb]
   \includegraphics[width=7.0cm]{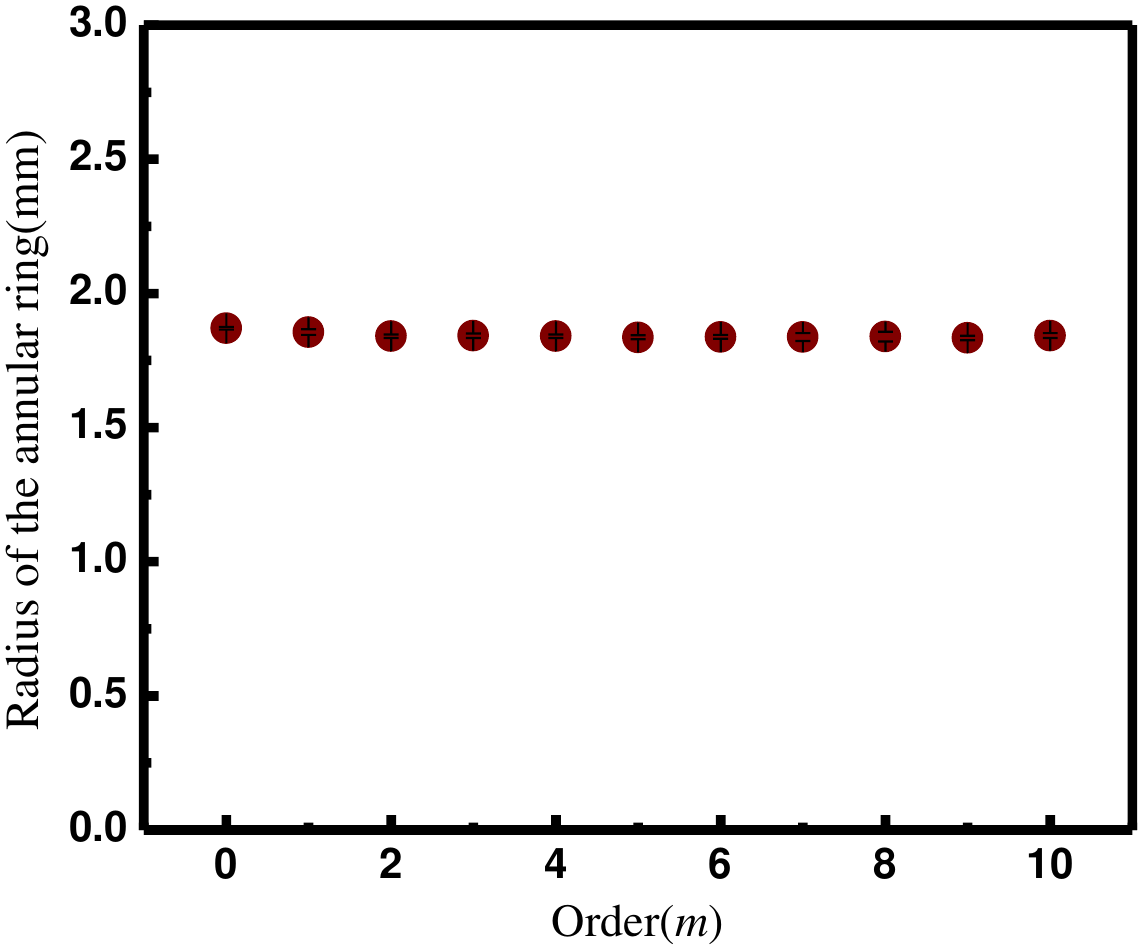}
    \caption{\textit{(Colour online) The variation of the radius of the annular ring with the order of experimentally generated POV beams.}}
    \label{fig:radius}
   \end{figure}
   \end{center} 

We start our experiment by recording the intensity distribution of POV beams at the Fourier plane of lens L1. For verifying the order independent intensity distribution, we have measured the radius of the bright ring for different orders of the POV beam, and the results are shown in Fig. \ref{fig:radius}. The radius has been calculated by averaging over 20 line profiles (i.e. 40 radii values) taken along diameters in different directions. From the figure, it is clear that the radius is independent of the order and confirms the quality of our POV beams. The average radius of the annular ring of the POV beams is obtained as  $\rho_0=1.844\pm 0.013 mm$, which has been used in all our theoretical calculations.

   \begin{center}
   \begin{figure}[htb]
   \includegraphics[width=7.0cm]{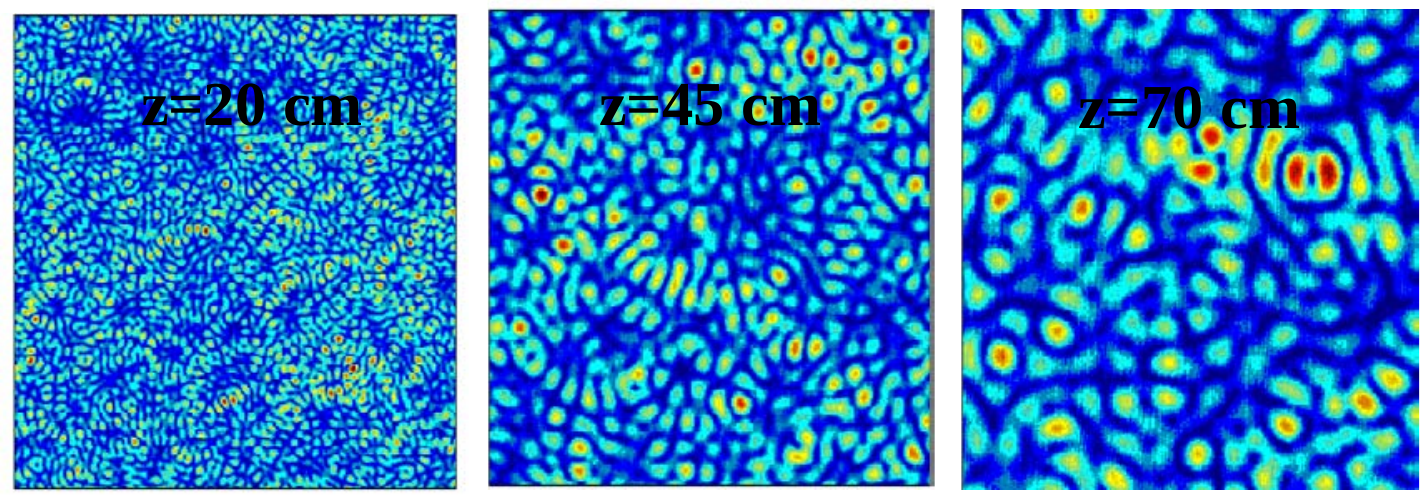}
    \caption{\textit{(Colour online)The speckle patterns obtained by the scattering of POV beam of order $m=2$ at different propagation distances z=20cm, z=45cm and z=70cm (from left to right) in the near field.}}
    \label{fig:expt2}
   \end{figure}
   \end{center}

We have recorded the speckle patterns corresponding to different orders of the POV beams by placing a camera immediately after the GGP. Fig \ref{fig:expt2} shows the speckle patterns obtained by scattering a second order POV beam and observed at different propagation distances $z = 20, 45,$ and $70 cm$ (from left to right). It is evident from the figure that the size of the speckles increases with propagation distance, i.e. the near-field speckles are diverging in nature.   

 \begin{center}
   \begin{figure}[htb]
   \includegraphics[width=7.0cm]{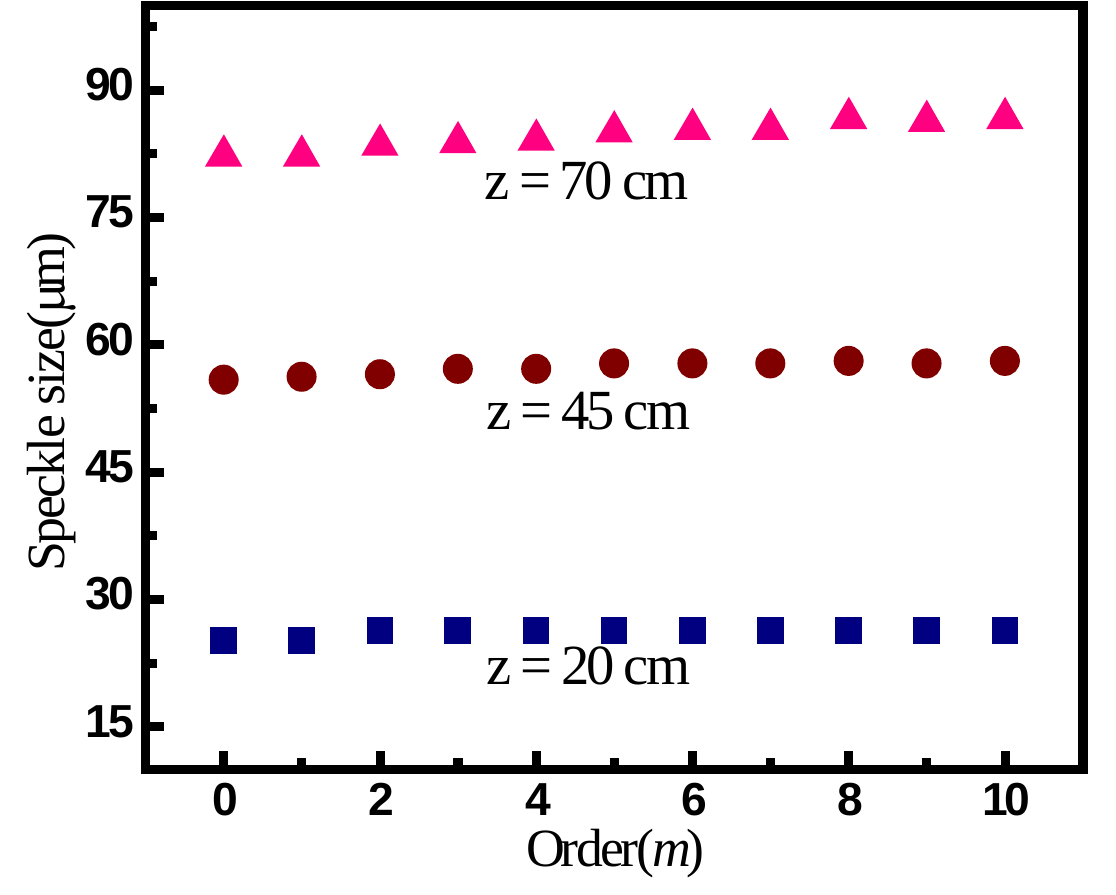}
    \caption{\textit{(Colour online) Variation of speckle size with the order of POV beam for different propagation distances $z=20cm$ (blue squares), $z=45cm$ (red circles), $z=70cm$ (pink triangles).}}
    \label{fig:Spec Size1}
   \end{figure}
   \end{center}
   
We have also studied the variation of speckle size, obtained as the width of auto-correlation function, with order and found that the size of the speckles is independent of the order of the POV beam. It has been shown graphically in Fig. \ref{fig:Spec Size1}. We have averaged the speckle size at a given propagation distance over 10 samples for each of the 11 orders $m = 0-10$.

  \begin{center}
   \begin{figure}[htb]
   \includegraphics[width=7.0cm]{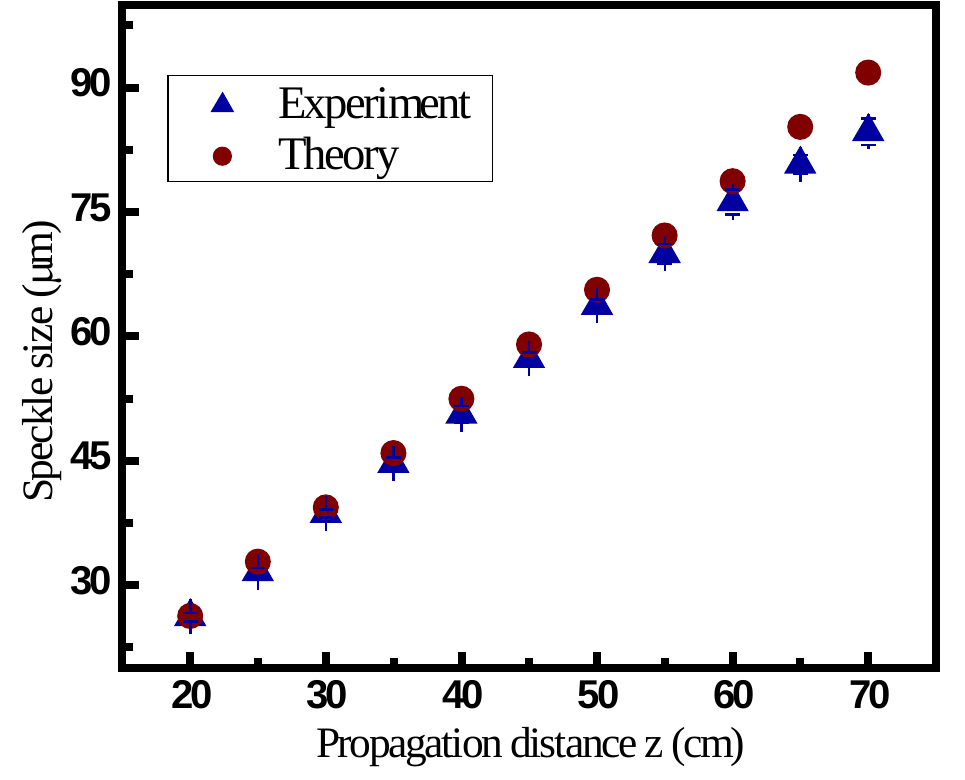}
    \caption{\textit{(Colour online) Experimental (blue triangles) and theoretical (red circles) results for the variation of near-field or diverging speckle size with propagation distance.}}
    \label{fig:Spec Size}
   \end{figure}
   \end{center}

\begin{center}
   \begin{figure}[htb]
   \includegraphics[width=7.0cm]{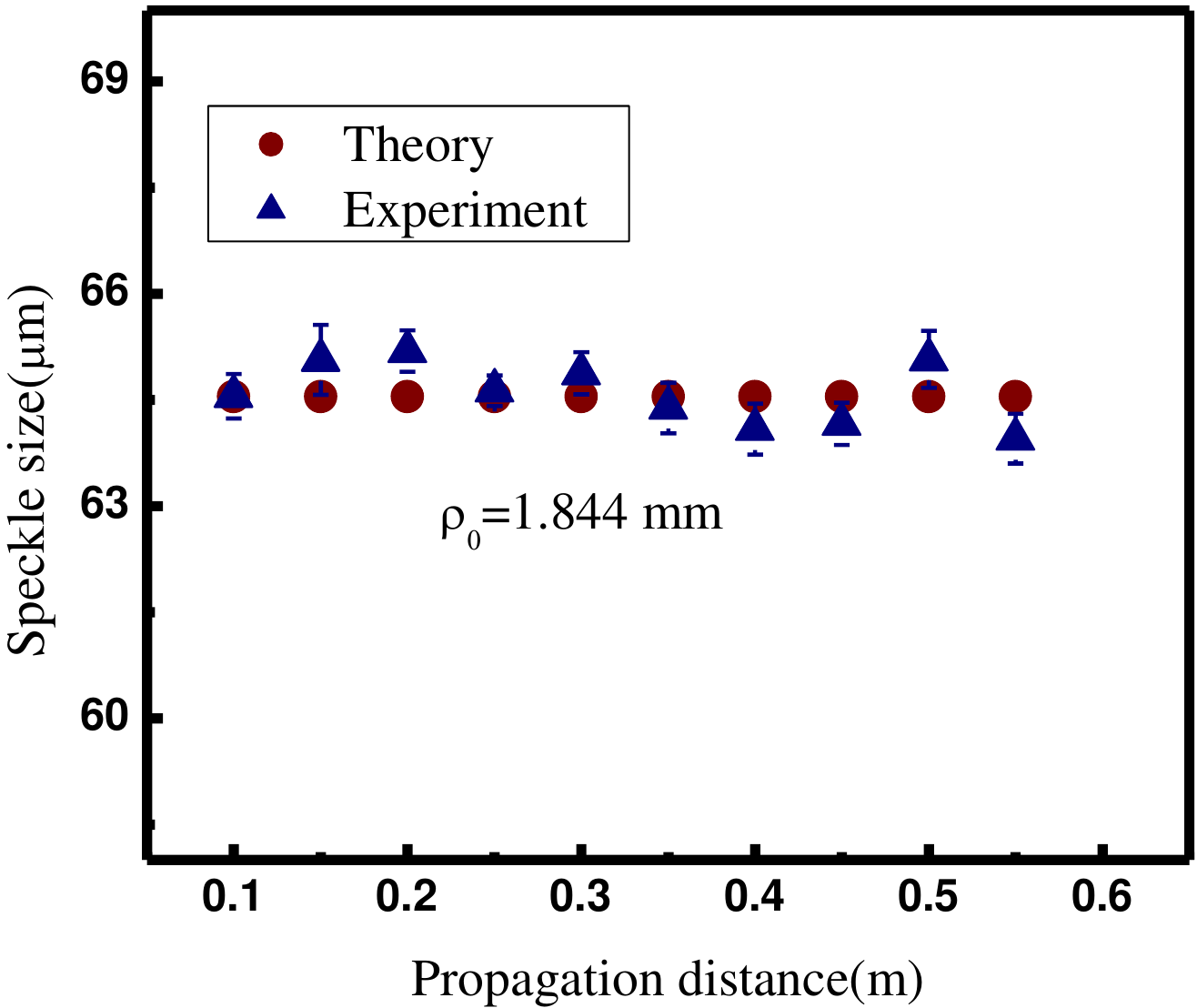}
    \caption{\textit{(Colour online) Experimental (blue triangles) and theoretical (red circles) results for the variation of speckle size with propagation distance z.}}
    \label{fig:spec farfield}
   \end{figure}
   \end{center}
  
  Figure \ref{fig:Spec Size} shows the theoretical (red circles) and experimental (blue triangles) results for the speckle size (now averaged over POV orders) at different propagation distances. From the figure, it is clear that the experimental results are well matching with theoretical values. For finding the theoretical speckle size, we have substituted the experimentally obtained ring radius $\rho_0 = 1.844\pm 0.013 mm$ in Eq.\ref{near field}.  From Fig. \ref{fig:Spec Size}, one can easily verify that the speckle size varies linearly with propagation distance, which is in agreement with our theoretical results.

 We have generated the non-diffracting random fields by taking the Fourier transform of near field speckles (speckle FT), by placing a lens (L2) so that the GGP is in the front focal plane of this lens. Fig. \ref{fig:spec farfield} shows the variation of the size of non-diffracting speckles with the propagation distance. It is evident from the figure that the speckle size is invariant with propagation distance and confirms the non-diffracting nature of speckles. Our experimental results are in good agreement with the theoretical calculations, not only in the sense that they are non-diffracting but quantitatively as well.

\section{Conclusions}

We have proposed a scheme for generating the diffracting as well as non-diffracting speckles by scattering the POV beams. We have also provided the exact analytical expression for near-field and far-field speckle size. It has been proven that the near field speckle size is directly proportional to the propagation distance whereas far-field speckle size is independent of propagation distance. In both cases, the speckle size is independent of the order of the POV beam. We have compared our theoretical results with experimental results and found that they are in good agreement with each other, not only qualitatively but quantitatively. These results may find applications in cryptography and medical optics as we have non-diffracting speckles whose size can be controlled easily by changing the focal length of the lens used to take the Fourier transform of BG beams.

The authors declare no conflicts of interest.

\section*{\textbf{Acknowledgement}}

This work was partly supported by JSPS KAKENHI Grant Number JP20H05888.
The author SGR acknowledges financial support from DST-SERB through start-up research grant (SRG/2019/000857).

\bibliography{sample}

\begin{thebibliography}{10}
\urlstyle{rm}
\expandafter\ifx\csname url\endcsname\relax
  \def\url#1{\texttt{#1}}\fi
\expandafter\ifx\csname urlprefix\endcsname\relax\def\urlprefix{URL }\fi
\expandafter\ifx\csname doiprefix\endcsname\relax\def\doiprefix{DOI: }\fi
\providecommand{\bibinfo}[2]{#2}
\providecommand{\eprint}[2][]{\url{#2}}

\bibitem{heckenberg1992generation}
\bibinfo{author}{Heckenberg, N.}, \bibinfo{author}{McDuff, R.},
  \bibinfo{author}{Smith, C.} \& \bibinfo{author}{White, A.}
\newblock \bibinfo{journal}{\bibinfo{title}{Generation of optical phase
  singularities by computer-generated holograms}}.
\newblock {\emph{\JournalTitle{Optics letters}}} \textbf{\bibinfo{volume}{17}},
  \bibinfo{pages}{221--223} (\bibinfo{year}{1992}).

\bibitem{torres2011twisted}
\bibinfo{author}{Torres, J.~P.} \& \bibinfo{author}{Torner, L.}
\newblock \emph{\bibinfo{title}{Twisted photons: applications of light with
  orbital angular momentum}} (\bibinfo{publisher}{John Wiley \& Sons},
  \bibinfo{year}{2011}).

\bibitem{franke2008advances}
\bibinfo{author}{Franke-Arnold, S.}, \bibinfo{author}{Allen, L.} \&
  \bibinfo{author}{Padgett, M.}
\newblock \bibinfo{journal}{\bibinfo{title}{Advances in optical angular
  momentum}}.
\newblock {\emph{\JournalTitle{Laser \& Photonics Reviews}}}
  \textbf{\bibinfo{volume}{2}}, \bibinfo{pages}{299--313}
  (\bibinfo{year}{2008}).

\bibitem{molina2007twisted}
\bibinfo{author}{Molina-Terriza, G.}, \bibinfo{author}{Torres, J.~P.} \&
  \bibinfo{author}{Torner, L.}
\newblock \bibinfo{journal}{\bibinfo{title}{Twisted photons}}.
\newblock {\emph{\JournalTitle{Nature physics}}} \textbf{\bibinfo{volume}{3}},
  \bibinfo{pages}{305--310} (\bibinfo{year}{2007}).

\bibitem{yao2011orbital}
\bibinfo{author}{Yao, A.~M.} \& \bibinfo{author}{Padgett, M.~J.}
\newblock \bibinfo{journal}{\bibinfo{title}{Orbital angular momentum: origins,
  behavior and applications}}.
\newblock {\emph{\JournalTitle{Advances in optics and photonics}}}
  \textbf{\bibinfo{volume}{3}}, \bibinfo{pages}{161--204}
  (\bibinfo{year}{2011}).

\bibitem{reddy2013experimental}
\bibinfo{author}{Reddy, S.~G.}, \bibinfo{author}{Kumar, A.},
  \bibinfo{author}{Prabhakar, S.} \& \bibinfo{author}{Singh, R.}
\newblock \bibinfo{journal}{\bibinfo{title}{Experimental generation of
  ring-shaped beams with random sources}}.
\newblock {\emph{\JournalTitle{Optics letters}}} \textbf{\bibinfo{volume}{38}},
  \bibinfo{pages}{4441--4444} (\bibinfo{year}{2013}).

\bibitem{ostrovsky2013generation}
\bibinfo{author}{Ostrovsky, A.~S.}, \bibinfo{author}{Rickenstorff-Parrao, C.}
  \& \bibinfo{author}{Arriz{\'o}n, V.}
\newblock \bibinfo{journal}{\bibinfo{title}{Generation of the “perfect”
  optical vortex using a liquid-crystal spatial light modulator}}.
\newblock {\emph{\JournalTitle{Optics letters}}} \textbf{\bibinfo{volume}{38}},
  \bibinfo{pages}{534--536} (\bibinfo{year}{2013}).

\bibitem{chen2013dynamics}
\bibinfo{author}{Chen, M.}, \bibinfo{author}{Mazilu, M.},
  \bibinfo{author}{Arita, Y.}, \bibinfo{author}{Wright, E.~M.} \&
  \bibinfo{author}{Dholakia, K.}
\newblock \bibinfo{journal}{\bibinfo{title}{Dynamics of microparticles trapped
  in a perfect vortex beam}}.
\newblock {\emph{\JournalTitle{Optics letters}}} \textbf{\bibinfo{volume}{38}},
  \bibinfo{pages}{4919--4922} (\bibinfo{year}{2013}).

\bibitem{vaity2015perfect}
\bibinfo{author}{Vaity, P.} \& \bibinfo{author}{Rusch, L.}
\newblock \bibinfo{journal}{\bibinfo{title}{Perfect vortex beam: Fourier
  transformation of a bessel beam}}.
\newblock {\emph{\JournalTitle{Optics letters}}} \textbf{\bibinfo{volume}{40}},
  \bibinfo{pages}{597--600} (\bibinfo{year}{2015}).

\bibitem{gori1987bessel}
\bibinfo{author}{Gori, F.}, \bibinfo{author}{Guattari, G.} \&
  \bibinfo{author}{Padovani, C.}
\newblock \bibinfo{journal}{\bibinfo{title}{Bessel-gauss beams}}.
\newblock {\emph{\JournalTitle{Optics communications}}}
  \textbf{\bibinfo{volume}{64}}, \bibinfo{pages}{491--495}
  (\bibinfo{year}{1987}).

\bibitem{reddy2014higher}
\bibinfo{author}{Reddy, S.~G.}, \bibinfo{author}{Prabhakar, S.},
  \bibinfo{author}{Kumar, A.}, \bibinfo{author}{Banerji, J.} \&
  \bibinfo{author}{Singh, R.}
\newblock \bibinfo{journal}{\bibinfo{title}{Higher order optical vortices and
  formation of speckles}}.
\newblock {\emph{\JournalTitle{Optics Letters}}} \textbf{\bibinfo{volume}{39}},
  \bibinfo{pages}{4364--4367} (\bibinfo{year}{2014}).

\bibitem{hu2020does}
\bibinfo{author}{Hu, X.-B.}, \bibinfo{author}{Dong, M.-X.},
  \bibinfo{author}{Zhu, Z.-H.}, \bibinfo{author}{Gao, W.} \&
  \bibinfo{author}{Rosales-Guzm{\'a}n, C.}
\newblock \bibinfo{journal}{\bibinfo{title}{Does the structure of light
  influence the speckle size?}}
\newblock {\emph{\JournalTitle{Scientific reports}}}
  \textbf{\bibinfo{volume}{10}}, \bibinfo{pages}{1--11} (\bibinfo{year}{2020}).

\bibitem{goodman2007speckle}
\bibinfo{author}{Goodman, J.~W.}
\newblock \emph{\bibinfo{title}{Speckle phenomena in optics: theory and
  applications}} (\bibinfo{publisher}{Roberts and Company Publishers},
  \bibinfo{year}{2007}).

\bibitem{dainty2013laser}
\bibinfo{author}{Dainty, J.~C.}
\newblock \emph{\bibinfo{title}{Laser speckle and related phenomena}},
  vol.~\bibinfo{volume}{9} (\bibinfo{publisher}{Springer science \& business
  Media}, \bibinfo{year}{2013}).

\bibitem{franccon2012laser}
\bibinfo{author}{Fran{\c{c}}on, M.}
\newblock \emph{\bibinfo{title}{Laser speckle and applications in optics}}
  (\bibinfo{publisher}{Elsevier}, \bibinfo{year}{2012}).

\bibitem{jacquot2012interferometry}
\bibinfo{author}{Jacquot, P.} \& \bibinfo{author}{Fournier, J.-M.}
\newblock \emph{\bibinfo{title}{Interferometry in speckle light: Theory and
  applications}} (\bibinfo{publisher}{Springer Science \& Business Media},
  \bibinfo{year}{2012}).

\bibitem{ricklin2002atmospheric}
\bibinfo{author}{Ricklin, J.~C.} \& \bibinfo{author}{Davidson, F.~M.}
\newblock \bibinfo{journal}{\bibinfo{title}{Atmospheric turbulence effects on a
  partially coherent gaussian beam: implications for free-space laser
  communication}}.
\newblock {\emph{\JournalTitle{JOSA A}}} \textbf{\bibinfo{volume}{19}},
  \bibinfo{pages}{1794--1802} (\bibinfo{year}{2002}).

\bibitem{schouten2003phase}
\bibinfo{author}{Schouten, H.~F.}, \bibinfo{author}{Gbur, G.},
  \bibinfo{author}{Visser, T.~D.} \& \bibinfo{author}{Wolf, E.}
\newblock \bibinfo{journal}{\bibinfo{title}{Phase singularities of the
  coherence functions in young’s interference pattern}}.
\newblock {\emph{\JournalTitle{Optics letters}}} \textbf{\bibinfo{volume}{28}},
  \bibinfo{pages}{968--970} (\bibinfo{year}{2003}).

\bibitem{kumar2019image}
\bibinfo{author}{Kumar, P.}, \bibinfo{author}{Fatima, A.} \&
  \bibinfo{author}{Nishchal, N.~K.}
\newblock \bibinfo{journal}{\bibinfo{title}{Image encryption using
  phase-encoded exclusive-or operations with incoherent illumination}}.
\newblock {\emph{\JournalTitle{Journal of Optics}}}
  \textbf{\bibinfo{volume}{21}}, \bibinfo{pages}{065701}
  (\bibinfo{year}{2019}).

\bibitem{heeman2019application}
\bibinfo{author}{Heeman, W.} \emph{et~al.}
\newblock \bibinfo{journal}{\bibinfo{title}{Application of laser speckle
  contrast imaging in laparoscopic surgery}}.
\newblock {\emph{\JournalTitle{Biomedical optics express}}}
  \textbf{\bibinfo{volume}{10}}, \bibinfo{pages}{2010--2019}
  (\bibinfo{year}{2019}).

\bibitem{sirohi1999speckle}
\bibinfo{author}{Sirohi, R.}
\newblock \bibinfo{title}{Speckle metrology: Some newer techniques and
  applications}.
\newblock In \emph{\bibinfo{booktitle}{International Trends in Optics and
  Photonics}}, \bibinfo{pages}{318--327} (\bibinfo{publisher}{Springer},
  \bibinfo{year}{1999}).

\bibitem{cheng2007simplified}
\bibinfo{author}{Cheng, H.} \& \bibinfo{author}{Duong, T.~Q.}
\newblock \bibinfo{journal}{\bibinfo{title}{Simplified laser-speckle-imaging
  analysis method and its application to retinal blood flow imaging}}.
\newblock {\emph{\JournalTitle{Optics letters}}} \textbf{\bibinfo{volume}{32}},
  \bibinfo{pages}{2188--2190} (\bibinfo{year}{2007}).

\bibitem{fercher1981flow}
\bibinfo{author}{Fercher, A.} \& \bibinfo{author}{Briers, J.~D.}
\newblock \bibinfo{journal}{\bibinfo{title}{Flow visualization by means of
  single-exposure speckle photography}}.
\newblock {\emph{\JournalTitle{Optics communications}}}
  \textbf{\bibinfo{volume}{37}}, \bibinfo{pages}{326--330}
  (\bibinfo{year}{1981}).

\bibitem{kermisch1975partially}
\bibinfo{author}{Kermisch, D.}
\newblock \bibinfo{journal}{\bibinfo{title}{Partially coherent image processing
  by laser scanning}}.
\newblock {\emph{\JournalTitle{JOSA}}} \textbf{\bibinfo{volume}{65}},
  \bibinfo{pages}{887--891} (\bibinfo{year}{1975}).

\bibitem{reddy2016non}
\bibinfo{author}{Reddy, S.~G.} \emph{et~al.}
\newblock \bibinfo{journal}{\bibinfo{title}{Non-diffracting speckles of a
  perfect vortex beam}}.
\newblock {\emph{\JournalTitle{Journal of Optics}}}
  \textbf{\bibinfo{volume}{18}}, \bibinfo{pages}{055602}
  (\bibinfo{year}{2016}).

\bibitem{cottrell2007nondiffracting}
\bibinfo{author}{Cottrell, D.~M.}, \bibinfo{author}{Craven, J.~M.} \&
  \bibinfo{author}{Davis, J.~A.}
\newblock \bibinfo{journal}{\bibinfo{title}{Nondiffracting random intensity
  patterns}}.
\newblock {\emph{\JournalTitle{Optics letters}}} \textbf{\bibinfo{volume}{32}},
  \bibinfo{pages}{298--300} (\bibinfo{year}{2007}).

\bibitem{dudley2012controlling}
\bibinfo{author}{Dudley, A.} \emph{et~al.}
\newblock \bibinfo{journal}{\bibinfo{title}{Controlling the evolution of
  nondiffracting speckle by complex amplitude modulation on a phase-only
  spatial light modulator}}.
\newblock {\emph{\JournalTitle{Optics Communications}}}
  \textbf{\bibinfo{volume}{285}}, \bibinfo{pages}{5--12}
  (\bibinfo{year}{2012}).

\bibitem{vanitha2021correlations}
\bibinfo{author}{Vanitha, P.} \emph{et~al.}
\newblock \bibinfo{journal}{\bibinfo{title}{Correlations in scattered perfect
  optical vortices}}.
\newblock {\emph{\JournalTitle{Journal of Optics}}}  (\bibinfo{year}{2021}).

\bibitem{goodman2015statistical}
\bibinfo{author}{Goodman, J.~W.}
\newblock \emph{\bibinfo{title}{Statistical optics}} (\bibinfo{publisher}{John
  Wiley \& Sons}, \bibinfo{year}{2015}).

\bibitem{acevedo2018non}
\bibinfo{author}{Acevedo, C.~H.} \& \bibinfo{author}{Dogariu, A.}
\newblock \bibinfo{journal}{\bibinfo{title}{Non-evolving spatial coherence
  function}}.
\newblock {\emph{\JournalTitle{Optics letters}}} \textbf{\bibinfo{volume}{43}},
  \bibinfo{pages}{5761--5764} (\bibinfo{year}{2018}).

\bibitem{zwillinger2007table}
\bibinfo{author}{Zwillinger, D.} \& \bibinfo{author}{Jeffrey, A.}
\newblock \emph{\bibinfo{title}{Table of integrals, series, and products}}
  (\bibinfo{publisher}{Elsevier}, \bibinfo{year}{2007}).

\bibitem{bezerra2020sorting}
\bibinfo{author}{Bezerra, D.~O.}, \bibinfo{author}{Amaral, J.~P.},
  \bibinfo{author}{Fonseca, E.~J.}, \bibinfo{author}{Alves, C.~R.} \&
  \bibinfo{author}{Jesus-Silva, A.~J.}
\newblock \bibinfo{journal}{\bibinfo{title}{Sorting of spatially incoherent
  optical vortex modes}}.
\newblock {\emph{\JournalTitle{Scientific reports}}}
  \textbf{\bibinfo{volume}{10}}, \bibinfo{pages}{1--7} (\bibinfo{year}{2020}).

\bibitem{alves2016using}
\bibinfo{author}{Alves, C.~R.}, \bibinfo{author}{Jesus-Silva, A.~J.} \&
  \bibinfo{author}{Fonseca, E.~J.}
\newblock \bibinfo{journal}{\bibinfo{title}{Using speckles to recover an image
  after its transmission through obstacles}}.
\newblock {\emph{\JournalTitle{Physical Review A}}}
  \textbf{\bibinfo{volume}{93}}, \bibinfo{pages}{043816}
  (\bibinfo{year}{2016}).

\end{thebibliography}

\end{document}